\documentclass[12pt]{article}
\usepackage{latexsym}
\usepackage{amssymb,amsfonts,amsmath}
\usepackage{graphicx} 
\usepackage{indentfirst}
\usepackage{bbm}
\usepackage{amssymb}
\usepackage{verbatim}
\usepackage{amsmath, amsthm,amssymb}
\usepackage{mathrsfs}
\usepackage{hyperref}
\usepackage{amsfonts}
\usepackage{dsfont}
\usepackage{cite}

\newcommand {\cA}{{\cal A}}

\def\a{\alpha}
\def\b{\beta}

\def\d{\delta}
\def\e{\epsilon}

\def\g{\gamma}

\def\l{\lambda}
\def\m{\mu}
\def\n{\nu}
\def\o{\omega}
\def\p{\pi}

\def\r{\rho}
\def\s{\sigma}

\def\x{\xi}

\def\D{\Delta}
\def\F{\Phi}
\def\J{\Psi}
\def\L{\Lambda}
\def\O{\Omega}
\def\P{\Pi}

\def\ri{{\rm i}}
\def\re{{\rm e}}

\newcommand{\ve}{\varepsilon}                            

\newcommand{\pa}{\partial}                           
\newcommand{\hf}{\frac12}

\newcommand{\vf}{\varphi}
\newcommand{\be}{\begin{equation}}
\newcommand{\ee}{\end{equation}}
\newcommand{\bea}{\begin{eqnarray}}
\newcommand{\eea}{\end{eqnarray}}
\newcommand{\non}{\nonumber}
\newcommand{\ba}{\begin{array}}
\newcommand{\ea}{\end{array}}

\newcommand{\cu}{\underline{c}}

\def\double #1{#1{\hbox{\kern-2pt $#1$}}}

\newcommand{\bsubeq}{\begin{subequations}}
\newcommand{\esubeq}{\end{subequations}}

\newcommand{\rd}{\mathrm d}
\begin{document}

\begin{center}
{\large \bf Quantum equivalence of the Freedman-Townsend model \\
and the principal chiral $\sigma$-model}\footnote{This work,  completed in August 1986, was one of the earliest applications of the Batalin-Vilkovisky formalism. It was accepted for publication in  Sov. J. Nucl. Phys. in 1987. 
It was subsequently withdrawn shortly before publication, 
after the authors had been informed by a colleague
that the same problem had already been solved elsewhere. Due to a limited access to 
the journals, at the time it was not possible  to verify this information, 
which in fact turned out  to be false. 
}
\end{center}

\begin{center}
{\bf I. L. Buchbinder and S. M. Kuzenko}\footnote{Tomsk State University} \\
\vspace{5mm}

\footnotesize{Tomsk State Pedagogical Institute}
~\\
\vspace{2mm}
\end{center}

\begin{abstract}
\baselineskip=14pt
The Freedman-Townsend model is quantized using the Batalin-Vilkovisky approach to Lagrangian quantization of gauge theories with linearly dependent generators. Path integral arguments are then applied to demonstrate the quantum equivalence of the Freedman-Townsend model to the principal chiral $\sigma$-model.
\end{abstract}

\renewcommand{\thefootnote}{\arabic{footnote}}

Several years ago, Freedman and Townsend \cite{FT} constructed a model for a self-interacting gauge antisymmetric tensor field with classical action 
\bea \label{FTmodel}
{\mathscr S}_{\rm FT} (B,H) &=& \hf \int \rd^4 x\, \Big\{ \hf \ve^{\m\n \r\s} B^a_{\m\n} F^a_{\r\s}
- \frac{1}{g^2} (H^a_\m)^2\Big\} ~,\\
&&F^a_{\r \s} = \pa_\r H^a_\s - \pa_\s H^a_\r + f^{abc} H^b_\r H^c_\s~. \non
\eea
Here $B^a_{\m\n}(x) $ is the second-rank antisymmetric tensor field, $H^a_\m (x)$ is the auxiliary vector field, and $f^{abc}$ are the structure constants of a Lie group $G$. 
At the classical level, theory \eqref{FTmodel} was shown in \cite{FT} to be equivalent to the principal chiral $\s$-model 
\bea
{\mathscr S}_\s= -\frac{1}{2g^2} \int \rd^4 x \, {\rm tr} \Big( \pa_\m U^{-1} \pa^\m U\Big) ~,
\label{2}
\eea
where $U (x) = \re^{\O(x)} \in G$.\footnote{Here $\O (x) = \O^a (x)T^a$, with $T^a$ the generators of the Lie algebra of $G$, $[T^a, T^b] = f^{abc} T^c$, normalized by ${\rm tr}(T^a T^b) =-\d^{ab}$. } Unlike the $\s$-model, the Freedman-Townsend theory possesses a nontrivial gauge invariance of the form 
\bea
\d B_{\m\n} = \nabla_{\m} \l_{\n}  - \nabla_{\n} \l_{\m}  \equiv \nabla_{[\m} \l_{\n]} ~, \qquad \d H_\m = 0 \quad \implies \quad 
\d {\mathscr S}_{\rm FT} =0~.
\label{3}
\eea
Here the fields and the gauge parameters are Lie-algebra-valued, 
in particular $B_{\m\n} = B^a_{\m\n} T^a$, and $\nabla_\m$ denotes a covariant derivative with  connection $H_\m$, specifically $\nabla_\m \o = \pa_\m \o + \big[ H_\m , \o \big]$, with $\o = \o^a T^a$. 

The Freedman-Townsend model \eqref{FTmodel} is a gauge theory with the property that  the standard Faddeev-Popov  scheme \cite{FP} is not applicable for its quantization. The same situation takes place for theories describing a gauge antisymmetric rank-$n$ tensor field in $d$ dimensions, with $n<d$, for the superfield $N=1$, $d=4$ supergravity (see e.g. \cite{GGRS}), as well as for some of those theories that have  recently attracted much interest, which are: the covariant Green-Schwarz superstring \cite{Green:1983wt} and Witten's string field theory \cite{Witten:1985cc}. 
All these theories belong to the class of reducible gauge theories with the property that  the generators of the gauge transformations $\d \vf^i = R^i{}_\a (\vf) \x^\a$, 
which leave the action ${\mathscr S}(\vf)$ invariant, $(\pa_r{\mathscr S}(\vf)/\pa\vf^i )R^i{}_\a (\vf) \equiv 0$, 
 are linearly dependent. This means that 
the matrix $R^i{}_\a (\vf)$  has nontrivial zero-eigenvalue eigenvectors $Z^\a{}_{\a_1} (\vf)$ at a stationary point $\vf_0$ of the action ${\mathscr S}(\vf)$, $R^i{}_\a Z^\a{}_{\a_1} \big|_{\vf_0} =0$.\footnote{Following the terminology of \cite{BV}, a gauge theory with linearly dependent generators is called a first-stage theory if the eigenvectors 
$Z^\a{}_{\a_1}$ are linearly independent with respect to the $\a_1$ index. In case that the matrix $Z^\a{}_{\a_1}$ also possesses nontrivial zero-eigenvalue eigenvectors $Z^{\a_1}{}_{\a_2}$
at the stationary point, $Z^{\a}{}_{\a_1}Z^{\a_1}{}_{\a_2}\big|_{\vf_0} =0$, and  the latter are linearly independent with respect to the  index $\a_2$, we are dealing with a second-stage theory, and so on and so forth. }
When quantizing reducible gauge theories, it is necessary to introduce ghosts for ghosts and to take into account the constraints imposed on the gauge fixing functions. In addition, the algebra of the generators of the gauge symmetries in reducible gauge theories may be open, as in the case of the Green-Schwarz superstring 
\cite{Green:1983wt}. Approaches to quantize theories with linearly dependent generators have been proposed by several authors \cite{BV, Schwarz, Siegel, Kimura:1980aw, Obukhov, Thierry-Mieg:1982eby, BK85}.
In our opinion, the most powerful is the Batalin-Vilkovisky method \cite{BV} since it allows one to derive covariant covariant Feynman rules in arbitrary gauge theories with open gauge algebra and linearly dependent generators. 

Quantization of the Freedman-Townsend model was first studied in 
\cite{Thierry-Mieg:1982eby} based on the generalized geometric quantization scheme for the Yang-Mills theory \cite{Baulieu:1981sb}. As applied to the Freedman-Townsend model, however, this scheme leads to some uncertainties related to the freedom in a choice of BRST transformations for the antisymmetric tensor field. As a result, the authors of \cite{Thierry-Mieg:1982eby} obtained an incorrect spectrum of ghosts which does not reproduce the well-known Abelian limit \cite{Schwarz, Siegel, Kimura:1980aw}.
Correct quantization of the theory  \eqref{FTmodel} has been carried out in a recent paper \cite{deAlwis:1987fr} by making use of a modified version of the Faddeev-Popov scheme. At the same time, a powerful method to quantize theories of the type \eqref{FTmodel} has existed since 1983 \cite{BV}.  Therefore it is of interest to carry out quantization of  the Freedman-Townsend model within the framework of \cite{BV}.

The Freedman-Townsend model is a reducible first-stage gauge theory. Indeed, the generators of gauge transformation \eqref{3} have the form 
\bea
R^i{}_\a (\vf) ~ \to ~ R^a_{\m\n}{}^{\g b} = \nabla_{[\m} \d_{\n]}{}^\g \,\d^{ab} \, \d^4(x-x_0)
\eea
and possess the following `eigenvectors'
\begin{align}
 Z^\a{}_{\a_1} ~\to ~ Z_\g{}^{b\, \cu} &= \nabla_\g \d^{b \cu} \d^4(x-x_0) ~, \non\\
R^a_{\m\n}{}^{\g b} Z_\g{}^{b \,\cu} &= F_{\m\n}^b f^{abc} \d^4(x-x_1)
=-\hf \ve_{\m\n \r\s} f^{abc} \d^4(x-x_1) \frac{\d {\mathscr S}_{\rm FT} }{\d B^b_{\r\s} (x)}~,
\label{5}
\end{align}
which become the nontrivial zero-eigenvalue eigenvectors on the equation of motion for
$B_{\r\s}$.\footnote{The variational derivative in \eqref{5} is defined by 
$\d {\mathscr S}_{\rm FT} = \hf \int \rd^4x \, \d B^b_{\r\s} \, {\d {\mathscr S}_{\rm FT} }/{\d B^b_{\r\s}}$.}  
The underlined index $\cu$ in \eqref{5} labels the eigenvectors. 

Now let us recall the key features of the Batalin-Vilkovisky method \cite{BV} of the quantization of first-stage gauge theories. Given a gauge theory with classical action 
${\mathscr S} (\vf^i)$, generators $R^i{}_\a(\vf)$ and eigenvector $Z^\a{}_{\a_1} (\vf) $, 
we associate with it a new `supersymmetric' theory $S( \F^A, \F^*_A)$.
Here the set of fields $\F^A$
includes the original fields $\vf^i$,  
while $\F^*_A$ are so-called antifields. The statistics of $\F^*_A$ is opposite to that of 
$\F^A$, $\e(\F^*_A) = \e(\F^A) + 1$. The action functional $S( \F^A, \F^*_A)$ is a proper solution of the master equation
\bea
\frac{\pa_r S}{\pa \F^A} \frac{\pa_l S}{\pa \F^*_A} =0~,
\label{6}
\eea
under the boundary condition $S( \F, \F^*) \big|_{\F^* =0}= {\mathscr S}(\vf)$.
Then, the partition function for the original gauge theory $S(\vf) $ may be defined in the form 
\bea
Z = \int D \F D \F^* \, \m \,
\d \Big[ \F^*_A - \frac{\pa \J}{\pa \F^A} \Big] \, \exp \Big[\frac{\ri }{\hbar} 
S(\F , \F^*) \Big]~.
\label{7}
\eea
Here $\m$ is the quantum integration measure, which can be omitted within some regularization procedures such as the dimensional regularization, provided $S(\F, \F^*)$ is a local functional. The fermionic functional $\J (\F)$ in \eqref{7} is called the gauge fermion. Its choice is arbitrary modulo the only condition that the path integral \eqref{7} has no residual gauge freedom. The partition function $Z$ proves to be independent of the gauge fermion \cite{BV}, $Z_\J = Z_{\J + \d \J}$.

As demonstrated in \cite{BV}, a proper solution of the master equation \eqref{6} 
always exists for the so-called ``minimal'' set of fields
\bea
\F_{\rm min} = \Big\{ \vf^i, C^\a , \eta^{\a_1} \Big\}, \qquad 
\e(C^\a) =\e_\a +1 ~, \quad \e ( \eta^{\a_1}) = \e_{\a_1}
\label{8}
\eea
as a power series in antifields, 
\bea
S(\F_{\rm min} , \F^*_{\rm min} ) &=& {\mathscr S} (\vf) + \vf^*_i R^i{}_\a (\vf) C^\a 
+C^*_\a \Big( Z^\a{}_{\a_1}(\vf) \eta^{\a_1}+ T^\a_{\b\g} (\vf) C^\g C^\b \Big) \non \\
&&+ \vf^*_i \vf^*_j \Big( K^{ji}_{\a_1}(\vf)  \eta^{\a_1} + E_{\a\b}^{ij} (\vf)  C^\b C^\a \Big) + \dots
\label{9}
\eea
In this expression, $T^\a_{\b\g}$ are the structure functions corresponding to the algebra
with generators $R^i_\a$, the coefficients $E_{\a\b}^{ij}$ describe the openness of the gauge algebra, and the quantities $ K^{ji}_{\a_1}$ express the condition of linear dependence of the generators off the mass shell \cite{BV}, 
\bea
R^i{}_\a Z^\a{}_{\a_1}  - 2 \frac{\pa_r {\mathscr S} }{\pa \vf^j } K^{ji}_{\a_1} (-1)^{\e_j} =0~.
\label{10}
\eea
The minimal set $\F_{\rm min}$ may be extended by an arbitrary number of field pairs 
\bea
\L_{\dots}~, \P_{\dots} ~, \qquad \e(\L) = \e(\P) +1 \non
\eea
and replace the action \eqref{9} with the functional
\bea
S(\F_{\rm min} , \F^*_{\rm min} ) + \L^{* \dots} \P_{\dots}~,
\non
\eea
which trivially depends on the new fields and antifields. The resulting action is a proper solution of the master equation in the new space of fields. This freedom is useful for constructing a gauge fermion. As shown in \cite{BV}, in the case of first-stage theories the minimal set of field \eqref{8} should be extended by the following pairs of fields: 
\bea
\bar C_\a , \p_\a~, \quad \e(\bar C_\a) = \e_\a +1~; \quad 
\bar \eta_{\a_1} , \p_{\a_1}~,\quad \e(\bar \eta_{\a_1} )= \e_{\a_1} ~; \quad 
 \eta'^{\a_1} , \p'^{\a_1} ~, \quad \e (\eta'^{\a_1} ) = \e_{\a_1} ~.
\label{11}
\eea
With this extended set of fields, the proper solution of the master equation is 
\bea
S(\F , \F^* ) = S(\F_{\rm min} , \F^*_{\rm min} ) + \bar C^{* \a} \p_{\a} 
+ \bar \eta^{* \a_1} \p_{\a_1} + \eta^{\prime *}_{\a_1}  \p'{}^{\a_1}~.
\eea

Let us turn to the quantization of the Freedman-Townsend model \eqref{FTmodel}.
The corresponding minimal set of fields \eqref{8} is 
\bea
\F_{\rm min} = \Big\{ B^a_{\m\n}, H^a_\m, C^a_\m, \eta^a \Big\}~, \qquad 
\e(C^a_\m) =1~, \quad \e(\eta^a) =0~.
\eea
The gauge transformations \eqref{3} are Abelian, and therefore the coefficients 
$T^\a_{\b\g}$ and $E^{ji}_{\a\b}$ in \eqref{9} vanish in the case under consideration. 
It follows from the relations \eqref{5} and \eqref{10} that $ K^{ji}_{\a_1}$ is given by 
\bea
K^{ij}_{\a_1} ~\to ~ K^a_{\m\n}{}^b_{\r \s}{}^c = -\hf \ve_{\m\n\r\s} f^{abc} \d^4(x_1-x) \d^4(x_2-x)~.
\eea
It may be shown that the remaining terms in \eqref{9}, denoted by the ellipsis, 
vanish for the model under consideration. Therefore, the proper solution of the master equation \eqref{6} in the minimal sector of fields is given by 
\bea
S(\F_{\rm min} , \F^*_{\rm min} ) &=& {\mathscr S}_{\rm FT} (B, H) \non \\
&&-  {\rm tr} \int \rd^4 x\left\{ \hf B^{*\m\n} \nabla_{[\m} C_{\n]} + C^{*\m }\nabla_\m \eta 
+\frac 14 \ve_{\m\n\r\s} \big\{ B^{*\m\n}, B^{*\r\s}\big\} \eta
\right\}~.
\label{15}
\eea

Variables \eqref{11} for the  the Freedman-Townsend model are 
\bea
\bar C_\m^a , \p_\m^a~, \quad \e(\bar C_\m^a) = 1~; \quad 
\bar \eta^a , \p^a~,\quad \e(\bar \eta^a )= 0 ~; \quad 
 \eta'^{a} , \p'^{a} ~, \quad \e (\eta'^{a} ) = 0 ~. 
 \non
\eea
Therefore, the final proper solution of the master equation takes the form
\bea
S(\F , \F^*) =
S(\F_{\rm min} , \F^*_{\rm min} )  -{\rm tr} \int \rd^4 x \Big\{ \bar C^{*\m} \p _\m
+ \bar \eta^* \p  + \eta'{}^* \p'
\Big\} ~,
\label{16}
\eea 
where $ S(\F_{\rm min} , \F^*_{\rm min} ) $ is given by \eqref{15}. The last step of the Batalin-Vilkovisky quantization procedure is construction of a useful gauge fermion 
$\J(\F^A)$. We recall that the only restriction on $\J(\F^A)$ is the non-degeneracy 
of the path integral \eqref{7}. Our choice of the gauge fermion for the model \eqref{FTmodel} is 
\bea
\J = -{\rm tr} \int \rd^4x \Big\{ \bar C^\m \nabla^\n B_{\n\m} + \bar \eta \nabla_\m C^\m 
+ \bar C^\m \nabla_\m \eta' +\frac{1}{2m} \bar C^\m \p_\m +\frac{1}{2n}( \bar \eta \p' - \p \eta')\Big\}~,
\label{17}
\eea
for some numerical coefficients $n,m$. Plugging the relations \eqref{15}, \eqref{16} and \eqref{17} in \eqref{7} leads to the partition function of the Freedman-Townsend model
\bea \label{18}
Z&=& \int D \F \exp \Big[\frac{\ri}{\hbar} S_{\rm Q} (\F) \Big]~,\non \\
S_{\rm Q}(\F) &=& {\mathscr S}_{\rm FT}  (B,H) -{\rm tr} \int \rd^4x \bigg\{ 
-\hf \nabla^{[\m} \bar C^{\n]} \nabla_{[\m} C_{\n]} - \nabla^\m \bar \eta \nabla_\m \eta \\
&&+ \frac 14 \ve^{\m\n\r\s} \big\{ \nabla_{[\m} \bar C_{\n]} , \nabla_{[\r} \bar C_{\s]} \big\}\eta  
+\big( \nabla^\n B_{\n\m} + \nabla_\m \eta' + \frac{1}{2m} \p_\m \big) \p^\m 
\non \\
&&+\big( \nabla_\m C^\m + \frac{1}{2n} \p' \big) \p 
- \big( \nabla_\m \bar C^\m +\frac{1}{2n} \p \big) \p' \bigg\}~. \non
\eea
Upon doing the path integral over the Lagrange multipliers $\p_\m$, $ \p$ and $\p'$, 
we obtain a new representation for the partition function
\begin{subequations} \label{19}
\begin{align}
Z&= \int D \vf \exp \Big[\frac{\ri}{\hbar} S_{\rm Q} (\F) \Big]~,\qquad
\vf = \big\{ B_{\m\n} , H_\m , \bar C_\m , C_\m , \bar \eta, \eta, \eta'\big\} ~, \non \\
S_{\rm Q}(\F) &= {\mathscr S}_{\rm FT}  (B,H) +S_{\rm ghost} + S_{\rm gauge} ~,\\
S_{\rm ghost}&= \hf {\rm tr} \int \rd^4x \bigg\{ 
 \nabla^{[\m} \bar C^{\n]} \nabla_{[\m} C_{\n]} + \nabla^\m \bar \eta \nabla_\m \eta 
+\hf \ve^{\m\n\r\s} \big\{ \nabla_{[\m} \bar C_{\n]} , \nabla_{[\r} \bar C_{\s]} \big\}\eta  
\bigg\}~, \\
S_{\rm gauge} &= \hf {\rm tr} \int \rd^4 x \bigg\{ m \Big( \nabla^\n B_{\n\m} \nabla_\r B^{\r\m} 
+ \nabla^\m \eta' \nabla_\m \eta' -  B^{\m \n} \big[F_{\m\n} , \eta' \big]\Big)
-2n \nabla^\m \bar C_\m \nabla^\n C_\n \bigg\}~.
\end{align}
\end{subequations}
 
 It is worth discussing the obtained expressions \eqref{18} and \eqref{19} for the partition function of the Freedman-Townsend model. First of all, the ghost action $S_{\rm ghost}$
 contains a term describing the cubic interaction of the vector and scalar ghosts. This 
agrees with the results of \cite{deAlwis:1987fr}. However, $S_{\rm gauge} $ contains 
the contribution $B^{\m \n} \big[F_{\m\n} , \eta' \big]$ which is absent in 
 \cite{deAlwis:1987fr}. This difference is explained by the fact that our work and 
  \cite{deAlwis:1987fr} make use of different gauge fixing conditions for the local symmetry \eqref{3}. Our gauge fixing is $\nabla^\n B_{\n\m} =0$, while the authors of 
   \cite{deAlwis:1987fr} used the gauge condition $\pa^\n B_{\n\m}=0$.  
   As was pointed out above, the partition function \eqref{7} is independent of any particular choice of the gauge fermion $\J$, that is of the gauge fixing condition. 
 Had we stared with the following gauge fermion 
 \bea
\J = -{\rm tr} \int \rd^4x \Big\{ \bar C^\m \pa^\n B_{\n\m} + \bar \eta \pa_\m C^\m 
+ \bar C^\m \pa_\m \eta' +\frac{1}{2m} \bar C^\m \p_\m +\frac{1}{2n}( \bar \eta \p' - \p \eta')\Big\}~,
\non
\eea
we would have resulted with an expression for the partition function that would be equivalent to that found in  \cite{deAlwis:1987fr}. However, our choice of the gauge fermion proves to be more useful for the study of quantum equivalence of the Freedman-Townsend model \eqref{FTmodel} to the principal chiral $\s$-model \eqref{2}.
 
 It is easy to check that  the expression \eqref{16} obeys, along with eq, \eqref{6}, the following equation 
 \bea 
 \D S = 0~, \qquad \D = \frac{\pa_r}{\pa \F^A} \, \frac{\pa_l}{\pa \F^*_A} ~.
 \non
 \eea
 On this ground the local measure $\m$  appearing in \eqref{7} may be chosen to be
 $\m=1$ \cite{Batalin:1983ar} for the Freedman-Townsend model. As a result, the effective action \eqref{18} is invariant under the following BRST transformations \cite{BV}
 \bea
 \d \F^A = \frac{\pa_l S(\F, \F^*)} {\pa \F^*_A}\Big|_{ \F^*_A = \frac{\pa \J}{\pa \F} }~,
 \non
 \eea
 or in the explicit form 
 \begin{subequations} \label{20}
 \bea
 \d B_{\m\n} &=& \nabla_{[\m} C_{\n]} + \ve_{\m\n \r\s} \Big[ \nabla^{ [\r} \bar C^{\s]} ,\eta\Big]~,\\
\d H_\m &=& \d \eta = \d \p_\m = \d \p =\d \p' =0~, \\
\d C_\m &=& \nabla_\m \eta ~, \qquad \d \bar C _\m = \p_\m ~, \\
\d\bar \eta &=& \p ~, \qquad \d \eta' = \p'~.
 \eea
 \end{subequations}
 These transformations coincide with those found in   \cite{deAlwis:1987fr} modulo the gauge conditions chosen. In particular, the transformations \eqref{20} are nilpotent only on the mass-shell. 
 
 Now we turn to proving quantum equivalence of the Freedman-Townsend model \eqref{FTmodel} to the principal chiral $\s$-model \eqref{2}. For this we will show that the partition function of the principal chiral $\s$-model is obtained from \eqref{18}
 by performing a change of variables in the path integral. As a preliminary step we will massage the right-hand side of \eqref{18}. First of all, we point out that the three-ghost term $\ve^{\m\n\r\s} \big\{ \nabla_{[\m} \bar C_{\n]} , \nabla_{[\r} \bar C_{\s]} \big\}\eta  $
 does not contribute to the path integral. Next, in the path integral \eqref{18} we set 
 $m \to \infty$, $n=1$ and integrate over the fields $B_{\m\n}$, $\eta'$, $\p$ and $\p'$.  
This leads to 
\bea
Z&=& \int D H_\m D\bar C_\m D C_\m D \bar \eta D \eta D \p_\m \,
\d \big[ \hf \ve^{\m\n\r\s} F_{\r\s} + \nabla^{[ \m} \p^{\n]} \big] \d \big[ \nabla^\m \p_\m\big] 
\label{21} \\
&& \times {\rm exp} \bigg\{ - \frac{\ri}{\hbar} {\rm tr} \int \rd^4 x
\Big[ - \frac{1}{2g^2} H^\m H_\m -\hf \nabla^{[\m} \bar C^{\n]} \nabla_{[\m} C_{\n]}
+ \nabla^\m \bar C_\m \nabla^\n C_\n + \nabla^\m \bar \eta \nabla_\m \eta
\Big] \bigg\}~. 
\non 
\eea
 The following identity holds
 \bea
&&  \int D H_\m D \p_\m \, \cA (H_\m, \p_\m) \, \d \Big[ \hf \ve^{\m\n\r\s} F_{\r\s} + \nabla^{[ \m} \p^{\n]} \Big] \non \\
&=& \int D \O D \L \, \cA \Big( \bar H_\m (\O) , \bar \p_\m (\O, \L )\Big) 
{\det}^{-1/2} (\bar \square_2) \det (\bar \square_0)~.
 \label{22}
 \eea
 Here $\cA (H_\m, \p_\m)$ is an arbitrary functional, 
 \bea
 \bar H_\m (\O) = \re^{-\O} \pa_\m \re^{\O} ~, \quad \O = \O^a T^a~; \qquad 
 \bar \p_\m (\O, \L) = \bar \nabla_\m \L ~, \quad \L= \L^aT^a~.
 \non
 \eea 
 The covariant derivative $\bar \nabla_\m$ makes use of the connection $\bar H_\m$, 
 $\bar \square_0$ is the d'Alembertian of a scalar field $\vf^a$ in the adjoint representation of the group $G$, 
 \bea
 \bar \square_0 \vf^a = (\bar \nabla^\m \bar \nabla_\m \vf)^a ~. \non 
 \eea
  The d'Alembertian ${\bar \square}_2$  acts on the space of second-rank antisymmetric tensor fields in the adjoint
  representation, and is defined similarly. To prove the relation \eqref{22}, one should make use of the known lemma
\bea
\int \rd^n x_i\, \d^m \Big( f_\a (x_i) \Big) A (x) 
= \int \rd^{n-m} y_\l 
\Big[ \det \big( K_{\l\m} (y)\big) / \det \big( M_{\a\b} (y)\big) \Big]^{\hf} A\big(\bar x (y)\big) ~,  
\non 
\eea
 where $i=1, \dots, n$, $\a = 1, \dots, m$, $\l = 1, \dots n -m$, and $x^i = \bar x^i(y)$ is the solution of the equation $f_\a (x) =0$. The matrices $K_{\l\m}$ and $M_{\a\b}$ are defined as 
 \bea
 K_{\l\m} (y) = \Big( \frac{\pa \bar x^i}{\pa y^\l}  \frac{\pa \bar x^i}{\pa y^\m} \Big)\Big|_{x =\bar x(y)} ~,
 \qquad 
 M_{\a\b} (y) = \Big( \frac{\pa f_\a}{\pa x^i} \frac{\pa f_\b}{\pa x^i} \Big) \Big|_{x =\bar x(y)}
 ~. \non 
 \eea
 This lemma was used in \cite{Fradkin:1984ai}.
 
 Making use of \eqref{22}, the path integral \eqref{21} turns into
 \bea
 Z&=& \int D \bar C DC D \bar \eta D\eta D\O D\L\,\d \big( \bar \square \L\big) 
 {\det}^{-1/2} (\bar \square_2) \det ( \bar \square_0) \non \\
 && 
 \times {\rm exp} \bigg\{  \frac{\ri}{\hbar} {\rm tr} \int \rd^4 x
\Big[  \frac{1}{2g^2} \bar H^\m (\O) \bar H_\m(\O)  + \bar C^\m  \bar \nabla^\n \bar \nabla_\n C_\m
+\bar \eta \bar \nabla^\m  \bar \nabla_\m \eta
\Big] \bigg\}~. 
\eea
Doing the integration over $\bar C$, $C$, $\bar \eta$, $\eta$ and $\L$ gives
\bea
Z = \int D \O \,  {\det}^{-1/2} (\bar  \square_2) \det  ( \bar \square_1){\det}^{-1}  ( \bar \square_0)
{\rm exp} \bigg\{  \frac{\ri}{\hbar} {\rm tr} \int \rd^4 x
 \frac{1}{2g^2} \Big( \re^\O \pa_\m \re^{-\O}\Big)^2   \bigg\}~,
 \eea
 where  ${\bar \square}_1$  is the d'Alembertian on  the space of vector  fields in the adjoint. Since $\det  ( \bar \square_1) = {\det}^4  ( \bar \square_0)$ and 
 $\det  ( \bar \square_2) = {\det}^6  ( \bar \square_0)$, we observe that 
 \bea 
 {\det}^{-1/2} (\bar  \square_2) \det  ( \bar \square_1){\det}^{-1}  ( \bar \square_0)
=1~,
\non 
\eea
 and therefore
 \bea
Z = \int D \O \,  {\rm exp} \bigg\{  \frac{\ri}{\hbar} {\rm tr} \int \rd^4 x
 \frac{1}{2g^2} \Big( \re^{-\O} \pa_\m \re^{\O}\Big)^2   \bigg\}~.
 \eea
 Thus we have demonstrated that the partition functions of the Freedman-Townsend model \eqref{FTmodel} and the principal chiral $\s$-model \eqref{2} coincide. 
 This implies the equality of the ``on-shell'' effective actions in these theories
\cite{Fradkin:1984ai}.
 
 The above analysis will not change if we introduce a source term to the field $H_\m$ in 
 \eqref{18} by replacing 
 \bea
 {\mathscr S}_{\rm FT} (B,H) ~\to ~ {\mathscr S}_{\rm FT} (B,H) + \int \rd^4 x \, J^{\m a} H^a_\m~.
 \eea
 Repeating the same steps as above, we will then obtain formal equality of  the generating functionals in the Freedman-Townsend model 
 \bea
 Z(J) &=& \int D \F \exp \bigg\{ \frac{\ri} {\hbar} \Big[ S_{\rm Q} (\F) + \ \int \rd^4 x \, J^{\m a} H^a_\m \Big] \bigg\}
 \eea
 and in the principal chiral $\s$-model
 \bea
 Z_\s(J) = \int D \O \,  {\rm exp} \bigg\{  \frac{\ri}{\hbar} {\rm tr} \int \rd^4 x \Big[
 \frac{1}{2g^2} \Big( \re^{-\O} \pa_\m \re^{\O}\Big)^2  - J^\m  \re^{-\O} \pa_\m \re^{\O} \Big]\bigg\}~.
\eea


\begin{footnotesize}

\end{footnotesize}

\end{document}